\documentclass[10pt,twocolumn,letterpaper]{article}

\usepackage{wacv}
\usepackage{times}
\usepackage{epsfig}
\usepackage{graphicx}
\usepackage{amsmath}
\usepackage{amssymb}
\usepackage{subcaption}
\usepackage{breqn}

\def\wacvPaperID{854} 

\wacvfinalcopy 

\ifwacvfinal
\def\assignedStartPage{9876} 
\fi

\ifwacvfinal
\usepackage[breaklinks=true,bookmarks=false]{hyperref}
\else
\usepackage[pagebackref=true,breaklinks=true,colorlinks,bookmarks=false]{hyperref}
\fi

\ifwacvfinal
\else
\fi

\begin{document}

\title{Face Image Lighting Enhancement Using a 3D Model}

\author{Qiulin Chen\\
Purdue University\\
{\tt\small chen2114@purdue.edu}
\and
Jan P. Allebach\\
Purdue University\\
{\tt\small allebach@purdue.edu}
}

\maketitle

\begin{abstract}
  Image enhancement helps to generate balanced lighting distributions over faces.
  Our goal is to get an illuminance-balanced enhanced face image from a single view. Traditionally, image enhancement methods ignore the 3D geometry of the face or require a complicated multi-view geometry. Other methods cause color tone shifting or over saturation. Inspired by the new research achievements in face alignment and face 3D modeling, we propose an improved face image enhancement method by leveraging 3D face models. Given a face image as input, our method will first estimate its lighting distribution. Then we build an optimization process to refine the distribution. Finally, we generate an illuminance-balanced face image from a single view. Experiments on the FiveK dataset \cite{fivek} demonstrate that our method performs well and compares favorably with other methods.
  
\end{abstract}

\section{Introduction}


Illumination imbalance causes a lot of problems. Performance of models working on different tasks, such as face recognition, skin tone detection, and emotion recognition are greatly affected by imbalanced illumination. A lot of effort has been devoted to processing the photos taken under different illumination environments. Once a model can precisely describe the geometric relationship between the human face and the illumination environment, powerful portrait enhancement becomes applicable by manipulating the light distribution.
Image enhancement can help to balance the illumination distribution over face images.

Current enhancement methods can be divided into these five categories: 1) Histogram-based, 2) S-curve based, 3) Ratio-imaging-based, 4) Fusion-based, and 5) Learning-based methods. Histogram-based methods manipulate the histogram of the image directly to increase the image contrast, but fail when dealing with an underexposed image. S-curve methods apply pixel-level tone mapping to the original image. This increases the contrast in low lightness areas while compressing the value in high lightness areas. Bai and Reibman \cite{Bai2018Enhance} applied logarithmic mappings to the underexposure regions. They also built a model for underexposure and overexposure mapping detection. Ratio-imaging-based methods, to which our method belongs, separate the lighting distribution (shading) from the reflectance. These methods depend greatly on the accuracy of the lighting distribution estimation. Inaccurate estimation will result in ghosting artifacts, blurry restored images or uneven lightness. Fusion-based methods need images in multiple exposures and multiple s-curve mappings to fuse together. The final result is achieved by adjusting the illumination map and merging it with the reflectance. This will result in lower contrast and unnaturalness in the image. Learning-based methods perform well when dealing with small sets of exposure problems included in the training dataset, but they fail when dealing with images in the wild.

In this paper, we propose a new method to enhance face images. Inspired by the research on 3D face model and face alignment \cite{BlanzVolker1999Ammf} \cite{EggerBernhard20203MFM} \cite{Zhu20193DDFA} \cite{Zhou2020DPR}, we enhance the face image by utilizing a 3D face model estimated from a single portrait. Our approach first estimates an initial lighting distribution of an input face image. Then we design an optimization process to further improve the lighting distribution. Based on our prior knowledge, two regularizers are chosen. They focus on local illumination and global illumination, respectively. To increase the computational efficiency, we convert our convex optimization process into a linear system.

In Section \ref{sec:relate}, we introduce related work on image decomposition, spherical harmonics, face alignment, and a 3D face morphable model. Then, we introduce our method in Section \ref{faceEnhance}. In this section, we describe our method in three parts. In Section \ref{sec:compare}, we compare our model with others, both qualitatively and quantitatively. The experimental results demonstrate the effectiveness of our model for face image enhancement. 

The major contributions of this work are:
\begin{itemize}
    \item We apply a 3D face model to face image enhancement, which is rarely used in traditional face image enhancement methods.
    \item We design a loss function with two regularizers based on prior knowledge about the face lighting distribution. These two regularizers greatly improve the face enhancement results.
    \item To simplify the optimization process, we convert our convex optimization process into a linear system.
    \item Our model performs well and the experimental results show that our model compares favorable with other methods.
\end{itemize}

\section{Related Work } \label{sec:relate}
A lot of research has been done on face image enhancement. Image enhancement can help to improve the performance of models for different tasks such as face detection, face recognition, color constancy, and skin-tone detection. The method we use involves a lot of research from different areas including image enhancement, computer graphics, face alignment, and a 3D morphable model. 
The related work is introduced in the following order: 1) image decomposition, 2) spherical harmonics, 3) underexposed and overexposed image enhancement, 4) face alignment and a 3D face morphable model.

\subsection{Image Decomposition}
In computer graphics, researchers often treat an image as the multiplication between reflectance and irradiance \cite{Wen2003UIUCFaceRelight}. The irradiance is also known as the lighting distribution or shading. It describes how the lightness strength is distributed over the surface of the objects in the image. 

\begin{equation} \label{eq:intrinsicDecomp}
    I_p = S_pR_p + C_p
\end{equation}
In the above equation, Gross et al. \cite{Grosse2010IntrinsicImg} decomposed an image into three parts: shading (lighting distribution) $S_p$, reflectance $R_p$, and a specular term $C_p$. 
$S_pR_p$ represents the diffuse reflection, while $C_p$ represents specular reflection. $p$ represents the spatial coordinates. 
We decompose the images in the same way, except that we don't consider specular reflection independently.

\subsection{Spherical Harmonics} \label{SH}
Spherical harmonic functions provide a basis in frequency space to approximate functions distributed over spherical surfaces. 
For an arbitrary direction vector $\Vec{\omega}$, which is represented in a spherical coordinate system as in (\ref{eq:wvec}), the spherical harmonic basis functions \cite{JaroszDissertation2008} are defined in (\ref{eq:harmoicsFunc})-(\ref{eq:ledrenge-b}).

\begin{equation} \label{eq:wvec}
    \Vec{\omega} = (\sin{\theta} \cos{\phi}, \sin{\theta} \sin{\phi}, \cos{\theta})
\end{equation}

\begin{equation} \label{eq:harmoicsFunc}
    y_l^m(\theta, \phi)=\begin{cases}
      \sqrt{2}K^m_l \cos{(m\phi)}P^m_l(\cos{\phi}) & \text{if $m>0$,}\\
      K^0_lP^0_l(\cos{\theta}) & \text{if $m=0$,}\\
      \sqrt{2}K^m_l\sin{(-m\phi)}P^{-m}_l(\cos{\theta}) & \text{if $m<0$.}
    \end{cases}
\end{equation}

\begin{equation}
    K^m_l = \sqrt{\frac{(2l+1)}{(4\pi)} \frac{(l-|m|!)}{(l+|m|!)}}
\end{equation}


\begin{align} 
    &(P^0_0)_p = 1, \label{eq:ledrenge-a}\\
    &(P^m_m)_p = (2m-1)!!(1-p^2)^{m/2},\\
    &(P^m_{m+1})_p = \left(p(2m+1)P^m_m \right)_p,\\
    &(P^m_l)_p = \left(\frac{p(2l-1)}{l-m}P^m_{l-1}-\frac{(l+m-1)}{l-m}P^m_{l-2}\right)_p \label{eq:ledrenge-b}
\end{align}

In (\ref{eq:harmoicsFunc}), $l$ is the band index and $m$ represents the index in each band. $m$ satisfies the constraint that $-l<m<l$. $K^m_l$ are the normalization parameters while $P^m_l$ are the Associated Legendre Polynomials (ALP) \cite{DONG2002Ledrenge}. $!!$ represents a double factorial operation. A recurrence formula for ALP is provided from (\ref{eq:ledrenge-a}) to (\ref{eq:ledrenge-b}).

According to Ramamoorthi and Hanrahan \cite{Ramamorthi2001UCSD}, using the first nine basis terms is sufficient to approximate the irradiance map. 
The projection and combinations are calculated as in (\ref{eq:projection}) and (\ref{eq:combination}).

\begin{equation} \label{eq:projection}
    h_l^m = \int_{\Omega_{4\pi}}y^m_l(\Vec{\omega})S_{sphere}(\Vec{\omega})d\Vec{\omega}
\end{equation}
\begin{equation} \label{eq:combination}
    S_{sphere}^{\prime}(\Vec{\omega}) = \sum_{l=0}^{\infty}\sum_{m=-l}^{l}y^m_l(\Vec{\omega})h_l^m
\end{equation}

In (\ref{eq:projection}), $h_l^m$ represents the coefficient of the $y_l^m$ basis when projecting the light distribution $S_{sphere}(\Vec{\omega})$ to the basis functions. $l$ represents the band index, while $m$ represents the index in each band of spherical harmonics. $\omega$ represents the direction vector over the object surface. After getting all nine coefficients, we get the approximated light distribution $S_{sphere}^{\prime}(\Vec{\omega})$ over the object surface. We use a low-dimensional semidefinite programming (SDP) method to calculate the nine coefficients.

\subsection{Face alignment and 3D face morphable model} \label{morphable}
In (\ref{eq:combination}), we introduced a method to compute the illumination over a spherical surface. The basis functions are functions of the direction vector $\Vec{\omega}$. Usually, multi-view images of a single face are needed to precisely deduce the face normal. The 3D data collection process is pretty challenging. 
A lot of datasets were built in this way \cite{He2005Yale} \cite{sun2020SIPR} \cite{Debvec2001Acquire} \cite{Le2017UHDB31} \cite{Grosse2008Mutipie}. 
However, in real applications, multi-view photographs of a single person with fixed poses are usually not available. The face normal needs to be estimated from a single view. Face alignment and a 3D morphable model help to work around this problem \cite{BlanzVolker1999Ammf} \cite{EggerBernhard20203MFM}. 
Based on 3DMM, Zhu et al. \cite{Zhu20193DDFA} designed their 3D dense face alignment (3DDFA) model to tackle the challenges in face alignment. 
We leverage their idea to extract face normals. Then, we use the face normals to calculate the lighting distributions as is shown in (\ref{eq:combination}).

\section{Face Image Enhancement Using Face Normal} \label{faceEnhance}
Previously, we discussed the superiority of the spherical harmonics (SH) method in estimating the lighting distribution. We want to take advantage of this method to solve our problem focusing on face image enhancement. 
Our method can be divided into two parts. First, we use the spherical harmonics method to estimate an initial lighting distribution of the human face. This distribution is coarse and needs further refinement. Then we build our own quadratic loss function with two extra regularizers to optimize the lighting distribution. 

\subsection{Initialization of the lighting distribution}

There are many different ways to choose an initialization of the lighting distribution. As is shown in \cite{ZhangUnderExpo2018}, the maximum value across the red, green and blue (RGB) channels can be used as an initialization of the shading. However, directly using this to restore the image will result in a low contrast and unnatural image. Using the maximum value across the RGB channels will still contain a lot of high frequency components that are not only from lighting unevenness but also from reflectance. The enhanced image will have a low contrast problem. Instead, we use the estimation from the spherical harmonics (SH) as an initialization. The lighting distribution estimated from our method will be very smooth and only reflect the variations in illumination. 

\subsection{Refine the lighting distribution}
In order to refine the initial lighting distribution, 
we leverage the ideas from Xu et al. \cite{XuStructTV2014}. Our loss function is quadratic, and we can solve the optimization process by converting it into a linear system. Then, we can easily acquire a closed-form solution for it. This greatly reduces the computational complexity. 
Equation (\ref{eq:loss1}) shows that our loss function is composed of three parts. The first part comes from the original estimated lighting distribution. We want the global lightness order for both overexposure and underexposure regions to be similar to the original image. The second part controls the texture of the lighting distribution. We want the small texture components in the lighting distribution to be consistent with the original input image. This will help to reduce the artifacts in lighting unevenness and maintain the contrast level in the results. The third part controls the global illumination uniformity. We want to make the lightness order between overexposure and underexposure regions closer to each other. By doing this, we will get an enhanced face image with a more uniform lighting distribution. To achieve this, we utilize the face masks from the previous 3D morphable model and the overexposure and underexposure masks achieved by using the model from Bai et al. \cite{Bai2018Enhance}. This will help us to identify the overexposure and underexposure regions.

\begin{equation} \label{eq:loss1}
    s^{\prime} = \arg \min_{s} L(s) = L_{base}(s) + \lambda_g L_{g}(s) + \lambda_u L_{u}(s)
\end{equation}

In (\ref{eq:loss1}), $L_{base}(s)$ represents the constraint on the lighting distribution $s$ for lightness order similarity. $L_{g}(s)$ and $L_{u}(s)$ represent the constraints for local and global illumination uniformity in lighting distributions, respectively.

\textbf{Lightness order similarity.} The original images have overexposure and underexposure regions. During image enhancement, the lightness of underexposure regions may be mapped to be close to that of the overexposure regions to achieve a more uniformly illuminated face image. But the lightness order between the overexposure and underexposure regions should not be flipped. Violating this constraint may result in unnaturalness in the enhanced image. Equation (\ref{eq:loss-a}) shows the details of the first term. Constraining the refined lighting distribution map $s^{\prime}$ to be close to the original estimation will help retain the original lightness order. Furthermore, the values of the original estimated lighting distribution are truncated to ensure that when we restore the enhanced image from the lighting distribution according to $I_{enhanced} = I_{origin}/s^{\prime}$, the values will not be out of range. This constraint is similar to the color consistency constraint in \cite{ZhangUnderExpo2018}. But they use the maximum value across the RGB channels while we use the estimated lighting distribution from SH to be the initialization $s^{\prime}$. The advantage of our method is described previously. 

\begin{equation} \label{eq:loss-a}
    L_{base}(s) = \sum_{p}\left(s_p - s^{\prime}_p \right)^2
\end{equation}

\textbf{Local illumination uniformity.} After refining the lighting distribution according to the first and third terms in (\ref{eq:loss1}), the lighting estimation is still very coarse. The gradients of the illumination channel of the original image are included inside the second term in the loss function. This is to ensure that the relative texture strength in the enhanced image is consistent with the input image. Equations (\ref{eq:loss-b}) - (\ref{eq:loss-b-weight-2}) show details of the second term. There is a trade off in the estimated lighting distribution. If it contains too many texture components, the enhanced images will suffer from low contrast. If the distribution is too coarse with limited texture included, then the results may not achieve a uniform lighting distribution. Fine textures in the lighting distribution can bring more flexibility to the local lightness adjustment. In practice, a certain amount of texture information can help to remove the shadows cast by distant lighting sources on the side of the captured scene.

\begin{equation} \label{eq:loss-b}
    L_{g}(s) = \lambda_g \sum_p{\left( a_{x,p}\left( \frac{\partial s}{\partial x} \right)^2 + a_{y,p}\left( \frac{\partial s}{\partial y} \right)^2 \right)}
\end{equation}

\begin{align}
    a_{x,p} &= \left( \left|\frac{\partial Y}{\partial x} \right|^{\alpha} + \epsilon \right)^{-1}, \label{eq:loss-b-weight-1}\\
    a_{y,p} &= \left( \left|\frac{\partial Y}{\partial y} \right|^{\alpha} + \epsilon \right)^{-1}, \label{eq:loss-b-weight-2}
\end{align}

\begin{figure}
   \centering
    \subcaptionbox
      {Enhanced image from (b) with less texture.\label{enhanced-shadow1}}
      {\includegraphics[width=0.21\textwidth]{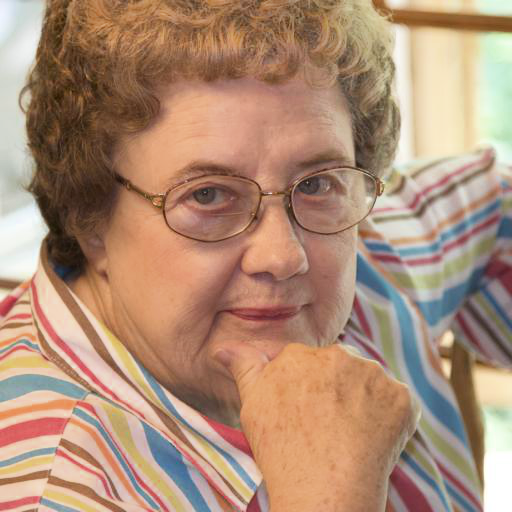}}
    \hskip 0.01truein
    \subcaptionbox
      {Lighting distribution with less texture.\label{lighting-shadow1}}
      {\includegraphics[width=0.21\textwidth]{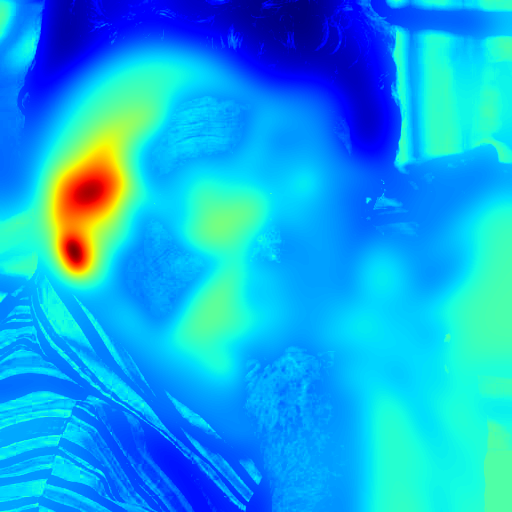}}
    \vskip 0.01truein
    \subcaptionbox
    {Enhanced image from (d) with more texture.\label{enhanced-shadow2}}
      {\includegraphics[width=0.21\textwidth]{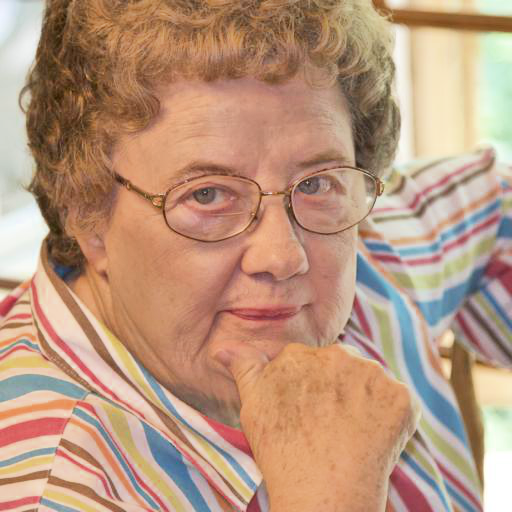}}
    \hskip 0.01truein
    \subcaptionbox
      {Lighting distribution with more texture.\label{lighting-shadow2}}
      {\includegraphics[width=0.21\textwidth]{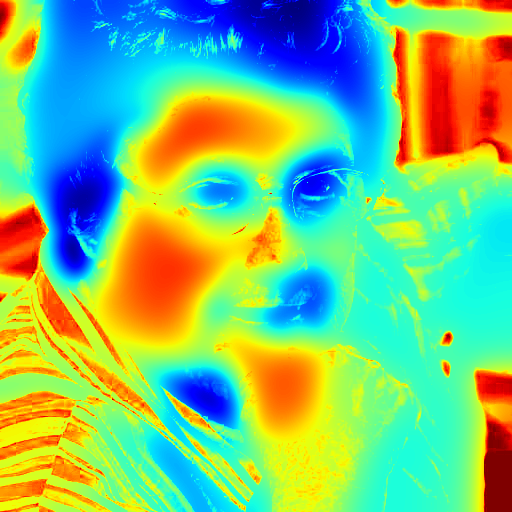}}
    \caption{Comparisons between two enhanced images with different amounts of texture in the lighting distribution.}
    \label{fig:textures}
\end{figure}

In (\ref{eq:loss-b}), the gradients of the lighting distribution $s$ are adjusted according to the gradients $\frac{\partial Y}{\partial x}$ and $\frac{\partial Y}{\partial y}$ of the luminance channels of the original image. Previously, a lot of research for skin detection worked successfully by using the orthogonal color space $YC_bC_r$ \cite{SHAIK2015Skintone, CHEDDAD2009ColorSkiin, Phung2005Skintone, Choudhury2008Skin, WONG2003faceColor}. The $Y$ channel is an additive combination of the RGB components. So it preserves the high frequency image content. Thus, we choose the $Y$ channel from the $YC_bC_r$ color space as the illumination. $\lambda_g$ in (\ref{eq:loss1}) is a parameter to control the weight of this term. $p$ represents pixel locations. The hyperparameter $\alpha$ is used to control the sensitivity to the gradients. Large gradients in the original illumination channel will result in small reciprocal values in the weights $a_{x, p}$ and $a_{y, p}$. This will slack the constraints on the gradients in the lighting distribution. So it will generate relatively large gradients at some spatial locations in the lighting distribution. Then, due to the relationship $I_{enhanced} = I_{origin}/s^{\prime}$, the locations in $s^{\prime}$ where the gradients are large can reduce the lightness nonuniformity in the original image. Thus, the enhanced image $I_{enhanced}$ will have an appearance much closer to that of photos taken under a uniform lighting environment. In Figure \ref{fig:textures}, we give an example.

Figure \ref{fig:textures} shows the different enhanced images when the lighting distribution $s$ contains different amounts of the texture information. 
Figure \ref{enhanced-shadow1} is an enhanced image from a relatively smooth lighting distribution while Figure \ref{enhanced-shadow2} is enhanced by retaining more texture in the lighting distribution. This is more obvious if we focus on the right eye regions of the subject in Figures \ref{lighting-shadow1} and \ref{lighting-shadow2}. Accordingly, the illumination around the right eye in the enhanced image in Figure \ref{enhanced-shadow2} is more uniform than that in Figure \ref{enhanced-shadow1}. 

\textbf{Global illumination uniformity.} In some cases, when a subject's face is lighted by a distant lighting source on one side of the scene, one side of the face will have very low illumination values while the other side has very large ones. To balance the two regions, we need an extra constraint. This is a more coarse balancing compared to the previous local uniformity constraint. Equation (\ref{eq:loss-c}) shows the details for this constraint.

\begin{equation} \label{eq:loss-c}
     L_{u}(s) = \lambda_u \sum_p{\left| \frac{M_u \odot I_p}{M_u \odot s_p} - \frac{M_o \odot I_p}{M_o \odot s_p} \right|^2}
\end{equation}

In (\ref{eq:loss-c}), $L_u$ represents the global uniformity constraint in the loss function. The parameter $\lambda_u$ controls its weight. $\odot$ represents element-wise multiplication. $s$ represents the lighting distribution. $I$ is the original input image. $p$ represents the pixel coordinates in input image. $M_u$ and $M_o$ are masks for the underexposure and overexposure regions, respectively \cite{Bai2018Enhance}, \cite{Guo2010Exposure}. This global illumination uniformity is to ensure that in the enhanced images, overexposure and underexposure regions will have exposure levels close to each other. The previous local illumination uniformity constraint will help to erase the shadows; but the differences between high exposure and low exposure regions will still exist. Our target is to make the lighting distributions within the overexposure masks be close to those within the underexposure masks. This is indicated by (\ref{eq:loss-c2}). So it is effectively equivalent if we take the reciprocal of the summand on both sides of the equation to get (\ref{eq:loss-c2-flip}). This is equivalent to minimizing the loss function shown in (\ref{eq:loss-c2-approx}). This approximation will simplify the optimization process.

\begin{equation} \label{eq:loss-c2}
     \frac{1}{N} \sum_p \left( \frac{I \odot M_u}{s \odot M_u}  \right) \approx \frac{1}{N} \sum_p \left(\frac{I \odot M_o}{s \odot M_o} \right)
\end{equation}

\begin{equation} \label{eq:loss-c2-flip}
     \frac{1}{N} \sum_p \left( \frac{s \odot M_u}{I \odot M_u} \right) \approx \frac{1}{N} \sum_p \left(\frac{s \odot M_o}{I \odot M_o} \right)
\end{equation}

\begin{equation} \label{eq:loss-c2-approx}
     L_{u} = \arg\min_{s} \frac{1}{N} \sum_p \left| \left( \frac{s \odot M_u}{I \odot M_u} \right)  - \left(\frac{s \odot M_o}{I \odot M_o} \right) \right |^2
\end{equation}

For the overexposure and underexposure masks, we leverage the ideas from Bai and Reibman \cite{Bai2018Enhance} and Guo et al. \cite{Guo2010Exposure}. Usually, we can deploy a hard-thresholding method to extract the likely overexposure and underexposure regions. So the pixel values larger than 254 will be assigned to the overexposure region and those less than 6 to 10 will be assigned to the underexposure regions. But this method does not deal well with the transition cases. And this will cause artifacts around the region boundaries. So a soft classification model is invented. Instead of doing hard thresholding, Guo et al. \cite{Guo2010Exposure} built a model to calculate the possibility for each pixel to be in overexposure class. They claimed that image pixels in the overexposure regions are desaturated and their lightness increases greatly. This results in an increase in $L^*$ channel and a decrease in the $a^*$ and $b^*$ channels. Inspired by this idea, Bai and Reibman created a similar model by comparing the relative values in the $L^*$, $a^*$ and $b^*$ channels to assign likelihood to pixels. Based on their idea, we designed our overexposure and underexposure mask region detector. Equation (\ref{eq:exposureMask}) shows the details of our model.

\begin{equation} \label{eq:exposureMask}
     \mathcal{L}(p)=\left(  \frac{\tanh(\delta(127+\alpha_T - ((G_{\sigma}*L^*)^2+\lVert C \rVert_2 )))+1}{ \tanh(\delta (127-\alpha_T+((G_{\sigma}*L^*)^2-\lVert C \rVert_2))) +1+\epsilon} \right)_p
\end{equation}

\begin{equation} \label{eq:ab_color}
    \lVert C \rVert_2 = (a^*)^2 + (b^*)^2\\
\end{equation}

\begin{align} \label{eq:exposureMask_ou}
    M_u &= \mathcal{L} >= 1 \\ 
    M_o &= \mathcal{L} < 1
\end{align}

In (\ref{eq:exposureMask}), $\sigma$ is a scale parameter to control the spread of the hyperbolic tangent function. $G_{\sigma}$ is the Gaussian filter kernel. $*$ represents the convolution operation. 
$\alpha_T$ is the parameter that controls the weight assigned to the overexposure and underexposure regions. When $\alpha_T$ is larger, our model is more aggressive in assigning pixels to the underexposure class. When it is smaller, the overexposure class assignment will be more aggressive. $\epsilon$ is a small coefficient. Finally, we achieve our binary mask for overexposure and underexposure regions by comparing the relative likelihood strength with 1. If larger than 1, we recognize the current pixel as one from the underexposure region, otherwise it is from the overexposure region. Empirically, we set the value of $\delta$ to be 1/60. For $\alpha_T$, we set it to be 128. 



\begin{figure}
   \centering
    \subcaptionbox
      {Enhanced image from (b).\label{fig:enhanced-shadow1}}
      {\includegraphics[width=0.21\textwidth]{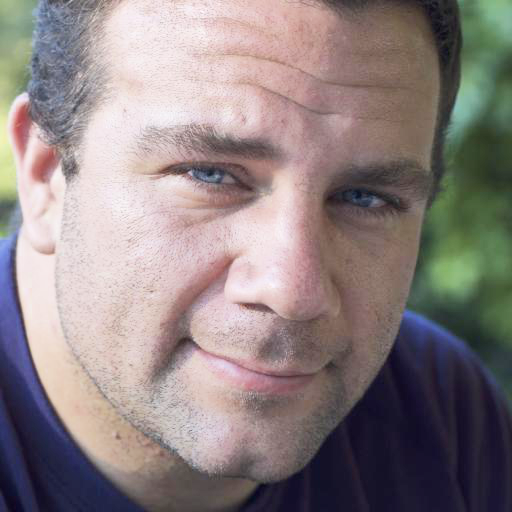}}
    \hskip 0.01truein
    \subcaptionbox
      {Lighting distribution with weak constraint.\label{fig:lighting-shadow1}}
      {\includegraphics[width=0.21\textwidth]{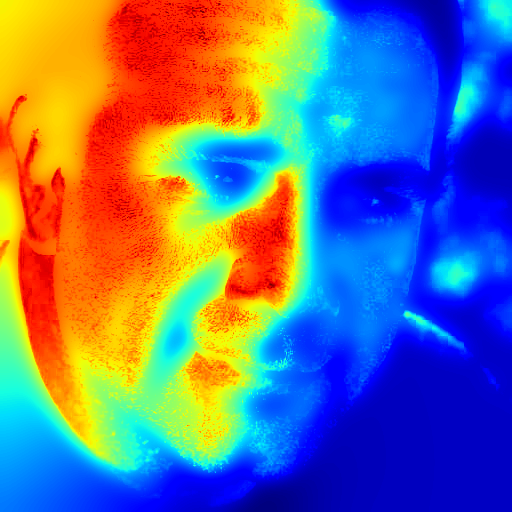}}
    \vskip 0.01truein
    \subcaptionbox
    {Enhanced image from (d).\label{fig:enhanced-shadow2}}
      {\includegraphics[width=0.21\textwidth]{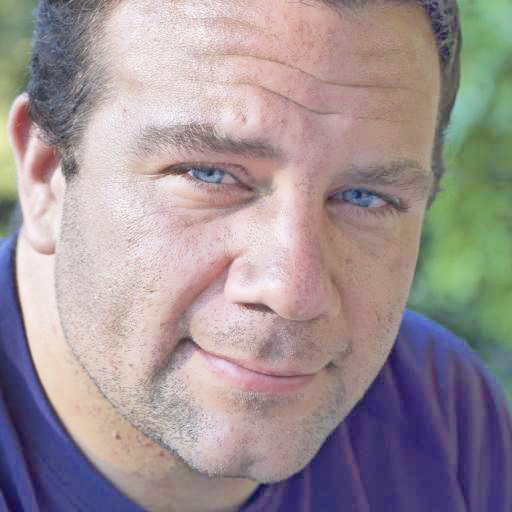}}
    \hskip 0.01truein
    \subcaptionbox
      {Lighting distribution with strong constraint.\label{fig:lighting-shadow2}}
      {\includegraphics[width=0.21\textwidth]{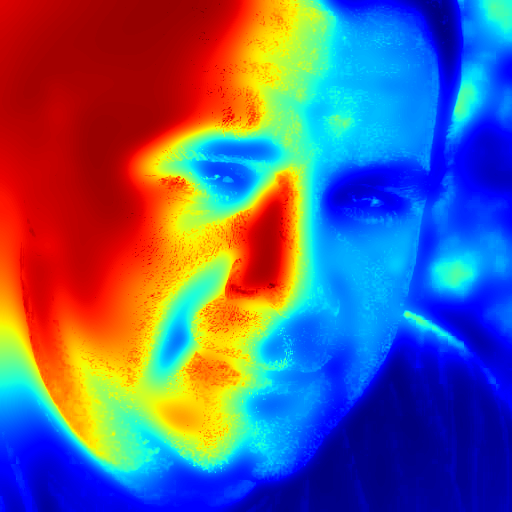}}
    \caption{Comparisons between two enhanced images with different degrees of global illumination uniformity constraint in the lighting distribution, as controlled by the paramter $\lambda_u$ in (\ref{eq:loss1}).}
    \label{fig:globalIllum}
\end{figure}

To further clarify the functionality of the global illumination uniformity, we show two more examples in Figure \ref{fig:globalIllum}. For the two enhanced images, by comparing the left and right half of the subject's face, we can see the lighting distribution on the subject's face is more uniform under the stronger constraint.

\subsection{Optimization solver}
Previously, we introduced the details in designing our loss function. We use approximations to both simplify the loss function and to achieve our goal to constrain the lighting distribution. Inspired by the Weighted Least Squares Method \cite{Farbman2008WLS} \cite{XuStructTV2014} \cite{ZhangUnderExpo2018}. Our loss function is a convex function and we can easily acquire a closed form solution for it. Due to our simplification, we can convert the quadratic loss function into a linear system. The process to achieve the linear system equations is shown below.

\begin{equation} \label{eq:lossAll}
    \begin{split}
    \arg \min_{s} L(s) = &\sum_{p}\left(s_p - s^{\prime}_p \right)^2\\ 
    + \lambda_g &\sum_p{\left( a_{x,p}\left( \frac{\partial s}{\partial x} \right)_p^2 + a_{y,p}\left( \frac{\partial s}{\partial y} \right)_p^2 \right)} \\ 
    + \lambda_u &\frac{1}{N} \sum_p \left| \left( \frac{s \odot M_u}{I \odot M_u} \right)_p  - \left(\frac{s \odot M_o}{I \odot M_o} \right)_p \right |^2
    \end{split}
\end{equation}

\begin{align} \label{eq:lossall-b-weight}
    a_{x,p} &= \left( \left|\frac{\partial Y}{\partial x}(p) \right|^{\alpha} + \epsilon \right)^{-1}, \\
    a_{y,p} &= \left( \left|\frac{\partial Y}{\partial y}(p)\right|^{\alpha} + \epsilon \right)^{-1}, \\
\end{align}

\begin{dmath} \label{eq:lossall-c-mask}
    \mathcal{L}(p) = \left(\frac{\tanh \left(\delta (127+\alpha_T - ((G_{\sigma} * L)^2 + \lVert C \rVert_2))\right) + 1}{\tanh \left(\delta (127-\alpha_T + ((G_{\sigma} * L)^2 - \lVert C \rVert_2))\right) + 1 + \epsilon} \right)
\end{dmath}

We have previously explained the variables in the above equations. Here, we just show how we achieve a simplified linear solver for this optimization problem. To simplify the expression, we use a high dimensional vector to represent the original image as $S \in R^{M\times N}$. $M$ and $N$ are the height and width, respectively, of the original input image. The loss function in this new format is shown in (\ref{eq:lossall-vec}) and (\ref{eq:lossall-K}).

\begin{equation} \label{eq:lossall-vec}
    \begin{split}
    \arg \min_{S} L(S) = &\left|S - S^{\prime} \right|^2\\ 
    + \lambda_g &{\left( S^T D^T_x A_x D_x S + S^T D^T_y A_y D_y S \right)} \\
    + \lambda_u & S^T K^T KS
    \end{split}
\end{equation}
\begin{equation} \label{eq:lossall-K}
    K = \left( [I+\epsilon]^{-1}_D [M_u]_D - [I+\epsilon]^{-1}_D [M_o]_D \right)
\end{equation}

In (\ref{eq:lossall-vec}), we show the loss function expressed in vector format. $S^\prime$ is the vector of the estimated lighting distribution. It is also used as the initialization value for $S$. The loss function is composed of a forward model and a two regularization terms weighted by $\lambda_g$ and $\lambda_u$. We use a grid search method to find the values of these two weights. $D_x$ and $D_y$ are the Toeplitz matrices \cite{Strang2014Toeplitz} for the forward difference. $A_x$ and $A_y$ are the inverse of the gradients of the original $Y$ channel. They are diagonal matrices. Elements on the diagonal are shown in (\ref{eq:lossall-b-weight}). $[\cdot]_D$ represents the diagonal form of the original matrix. $\epsilon$ is a small coefficient. $I$ is the original input image.

The loss function (\ref{eq:lossall-vec}) is a quadratic function in the variable $S$. So we can calculate the gradients of it in terms of the variable $S$. We can convert the quadratic convex optimization problem to a linear system. So the problem is reformatted as in (\ref{eq:leastSq}) - (\ref{eq:linearSys-H}).

\begin{dmath} \label{eq:leastSq}
    \frac{\partial L(S)}{\partial S} = (S-S^\prime)^T + \lambda_g \left(S^TD^T_xA_xD_x + S^TD^T_yA_yD_y \right) + \lambda_u S^T K^TK = 0
\end{dmath}

The double coefficients are merged into the weight parameters $\lambda_g$ and $\lambda_u$. $A_x$ and $A_y$ are diagonal matrices and they are symmetric.

\begin{equation} \label{eq:linearSys}
    \left( I_e + \lambda_g H^T + \lambda_u K^TK \right)S = S^\prime
\end{equation}

\begin{equation} \label{eq:linearSys-H}
    H^T = D^T_xA_xD_x + D^T_yA_yD_y
\end{equation}

$I_e$ is an identity matrix. $\left( I_e + \lambda_g H^T + \lambda_u K^TK \right)$ is the symmetric positive definite Laplacian matrix. The forward difference matrices are used to calculate the gradients along both $x$-axis and $y$-axis directions. And, by now, our loss function is in the format of $Ax=b$, a linear system, which is easy to solve. Also, we can use the sparsity of the matrices to accelerate the calculation process.

\section{Compare with Other Methods} \label{sec:compare}
\begin{figure}
  \centering
    {\includegraphics[width=0.4\textwidth]{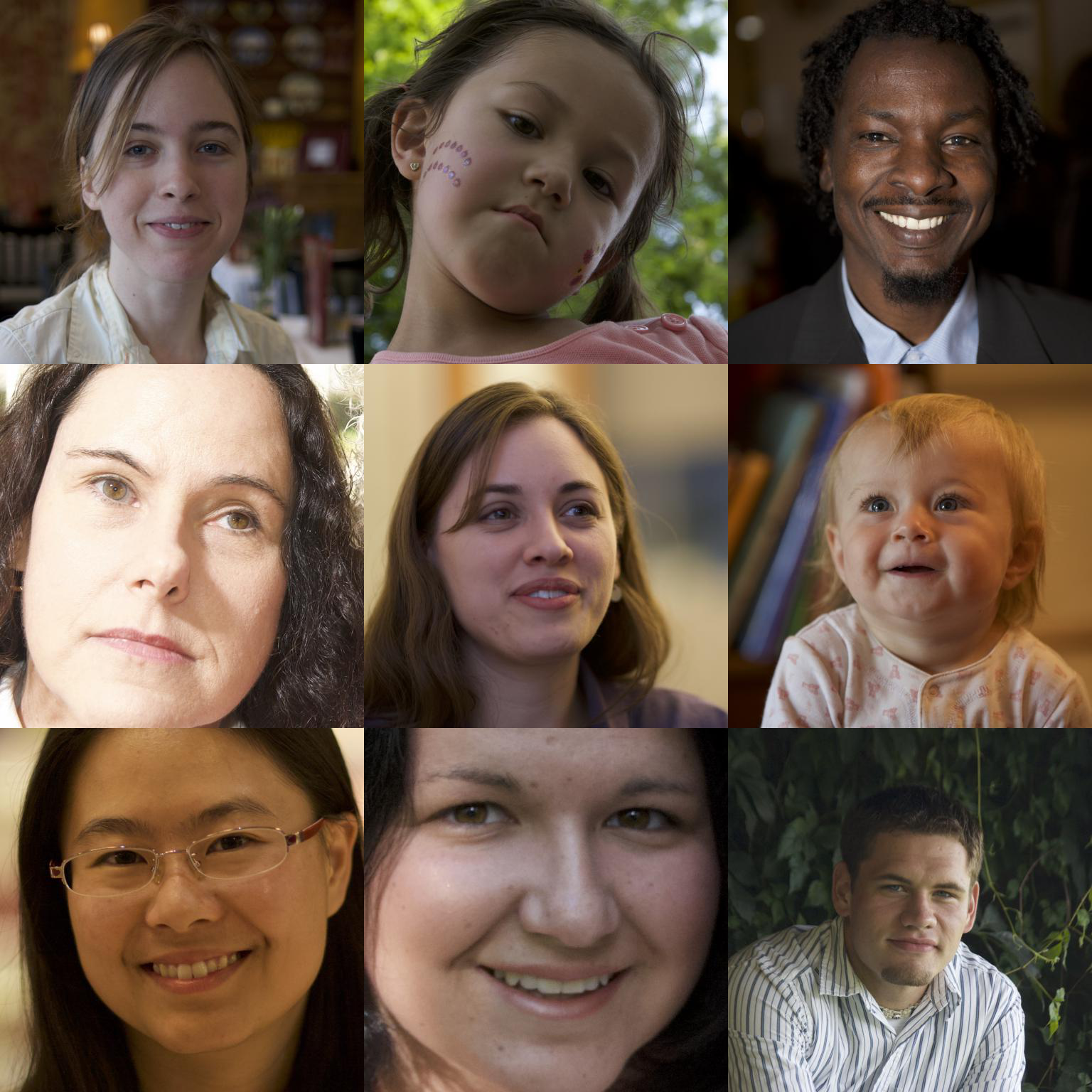}}
    \caption{Thumbnails of the nine testing images used to generate the results in Tables \ref{table:LOM} - \ref{table:DE}.}
    \label{fig:comparisons_montage9}
\end{figure}

To measure the performance of our method, underexposed portrait images from the MIT-Adobe FiveK Dataset \cite{fivek} are chosen to do a quantitative comparison between our method and other previous works. The thumbnails of the nine testing images are shown in Figure \ref{fig:comparisons_montage9}. We also choose one of these images for a qualitative comparison of the enhanced images.

\subsection{Qualitative comparisons}
\begin{figure}
  \centering
    \subcaptionbox
      {Original input image.\label{fig:inputImg5-1}}
      {\includegraphics[width=0.23\textwidth]{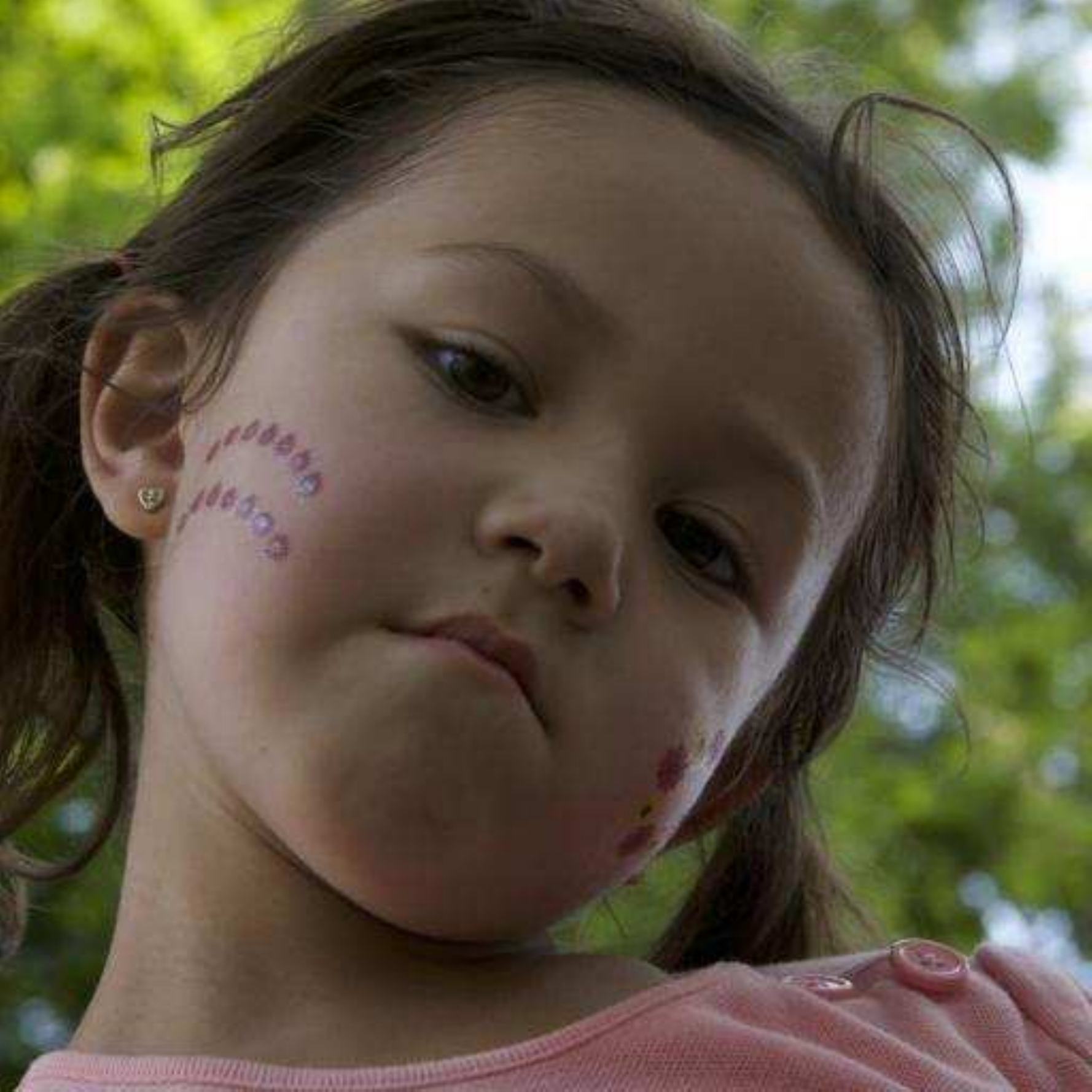}}
    \hskip 0.01truein
    \subcaptionbox
      {CLAHE \cite{PizerStephenM1987CLAHE}.\label{fig:CLAHE5}}
      {\includegraphics[width=0.23\textwidth]{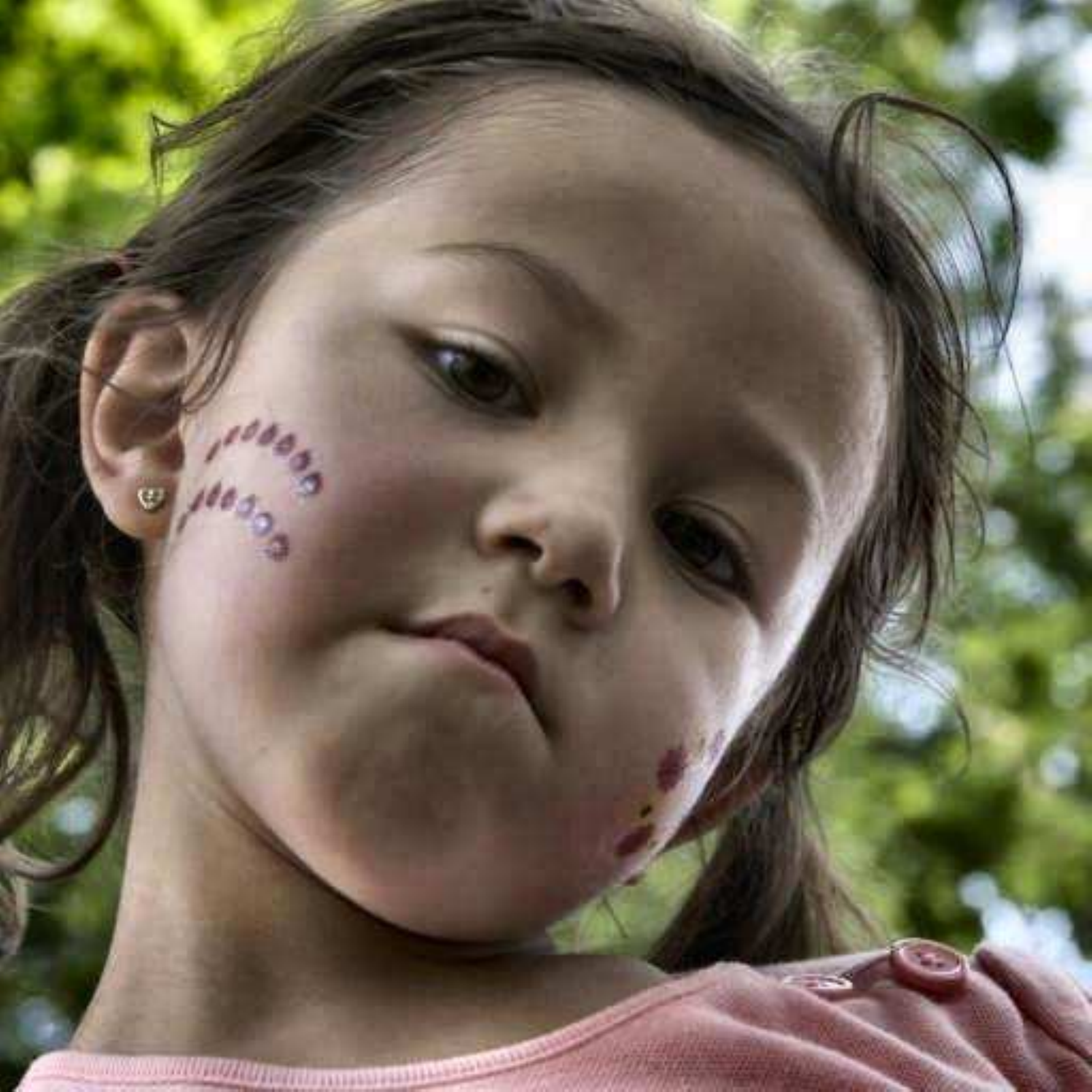}}
    \vskip 0.01truein
    \subcaptionbox
      {CIIE \cite{Bai2018Enhance}.\label{fig:CIE-Curve5}}
      {\includegraphics[width=0.23\textwidth]{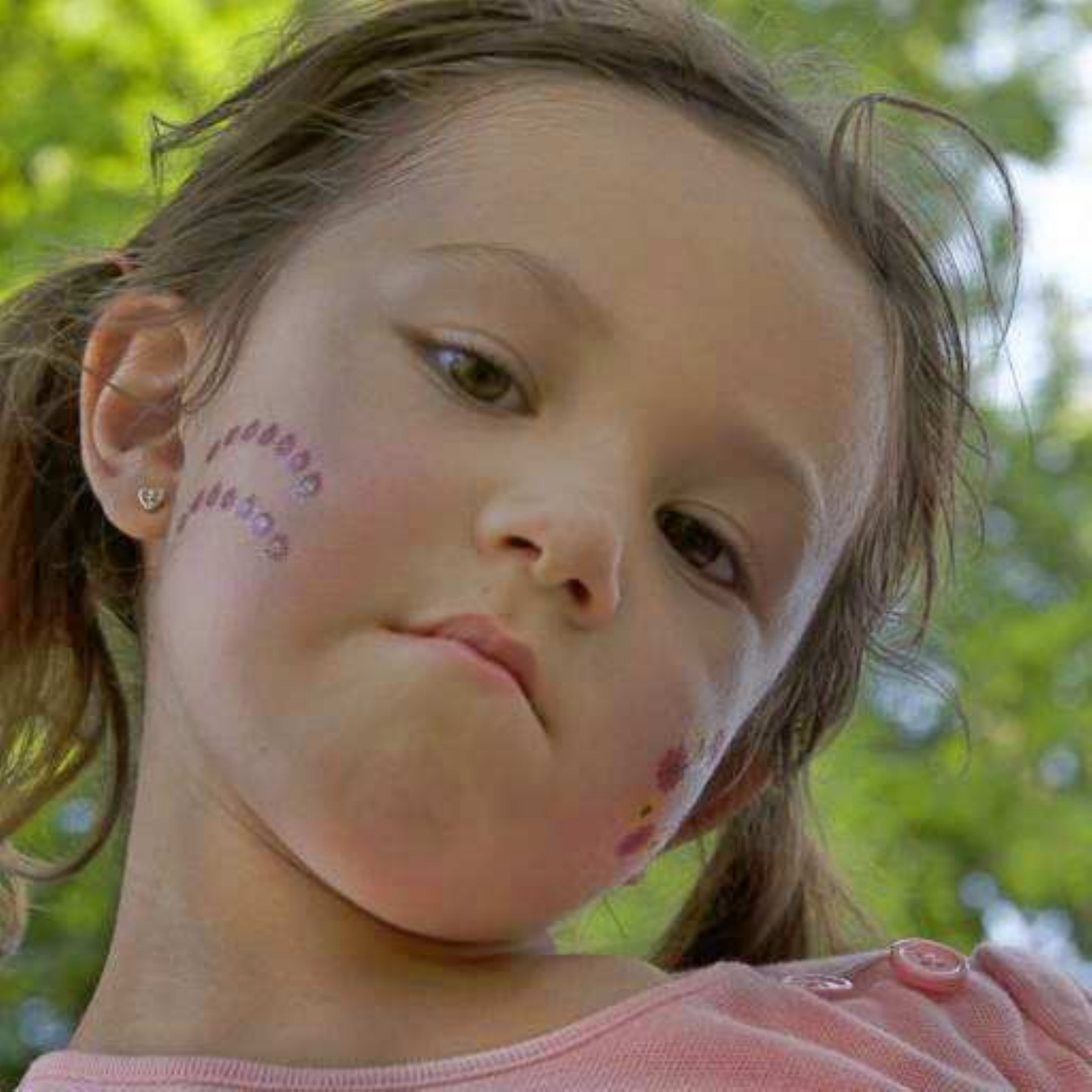}}
    \hskip 0.01truein
    \subcaptionbox
      {SRIE \cite{Fu2016SRIE}.\label{fig:SRIE5}}
      {\includegraphics[width=0.23\textwidth]{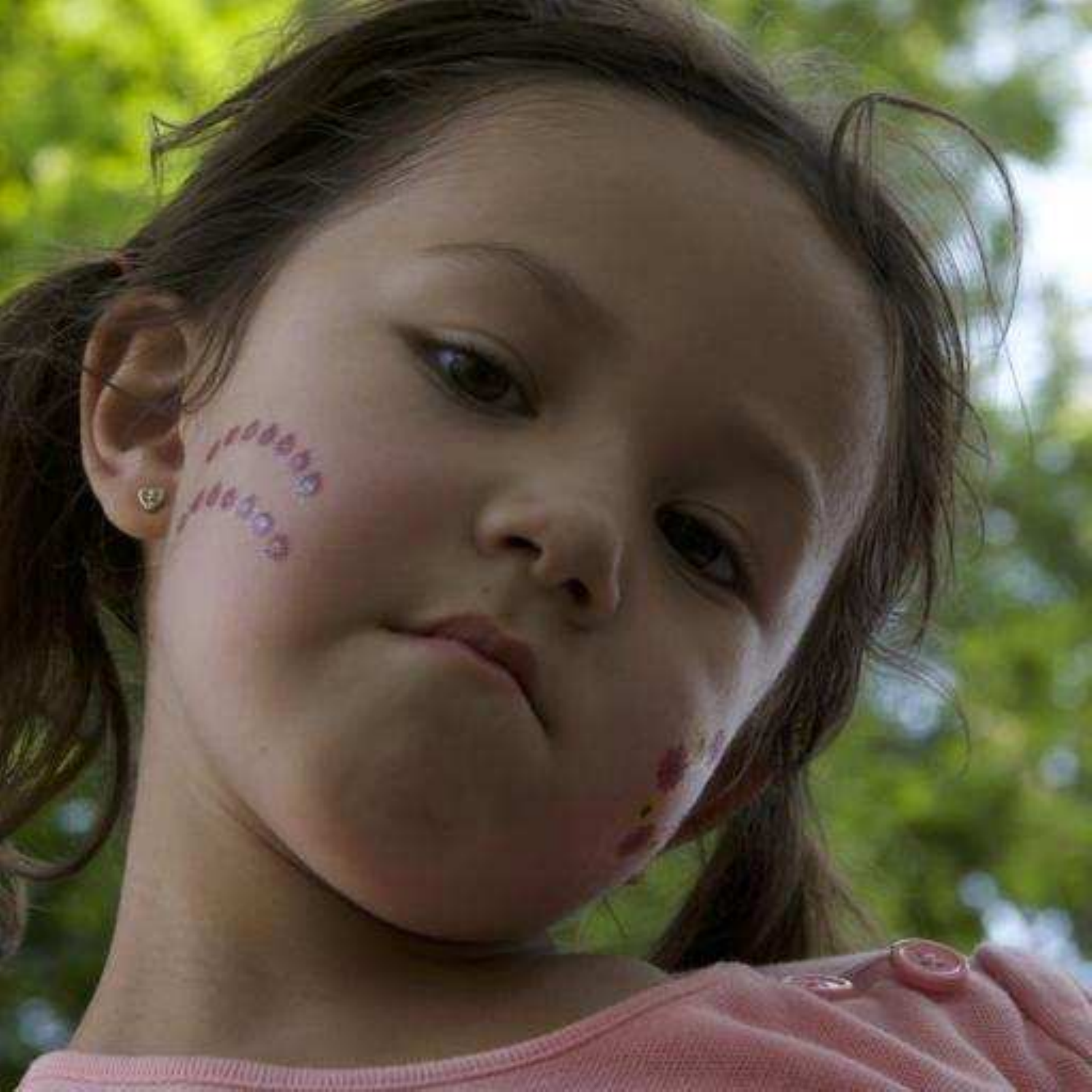}}
    \vskip 0.01truein
    \subcaptionbox
    {SRLL \cite{Li2018SRLL}.\label{fig:SRLL5}}
      {\includegraphics[width=0.23\textwidth]{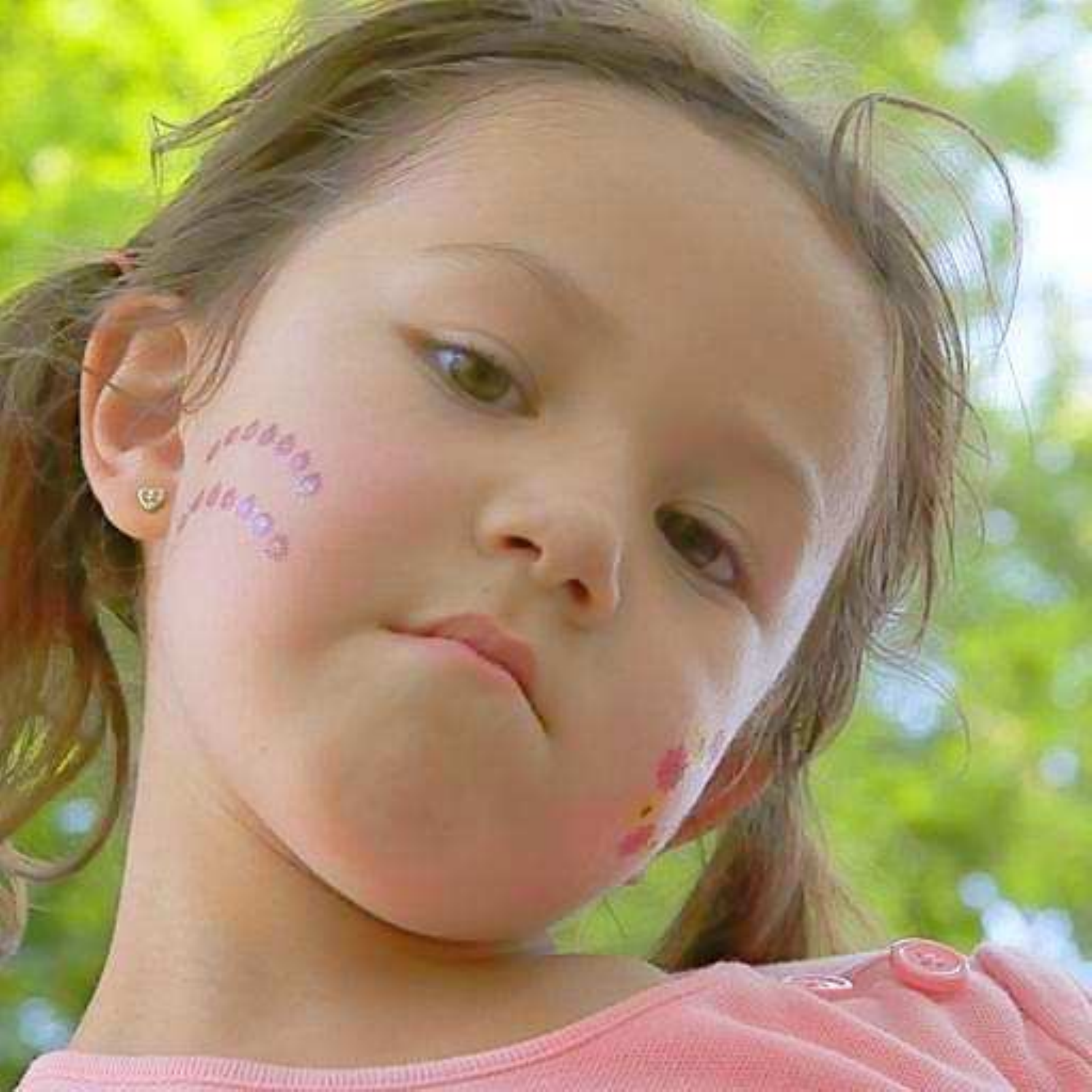}}
    \hskip 0.01truein
    \subcaptionbox
      {Ours.\label{fig:ours5}}
      {\includegraphics[width=0.23\textwidth]{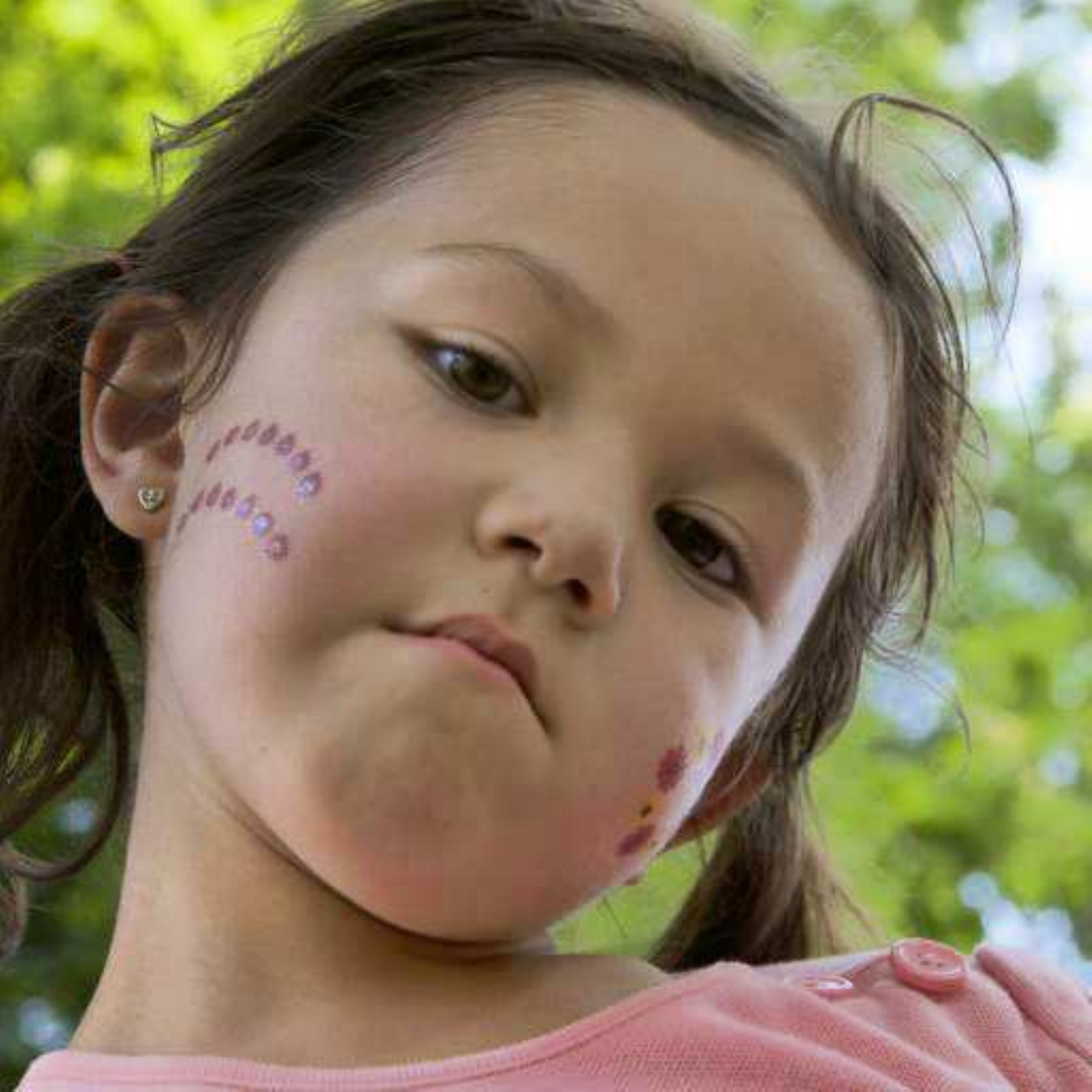}}
    \caption{Qualitative comparisons for different methods (CLAHE, CIIE, SRIE, SRLL, and ours).}
    \label{fig:comparisons2-5-1}
\end{figure}

\begin{figure}
  \centering
    \subcaptionbox
      {Original input image.\label{fig:inputImg5-2}}
      {\includegraphics[width=0.23\textwidth]{origin_5.pdf}}
    \hskip 0.01truein
    \subcaptionbox
      {LLCRM \cite{Ren2019LLCRM}.\label{fig:LLCRM5}}
      {\includegraphics[width=0.23\textwidth]{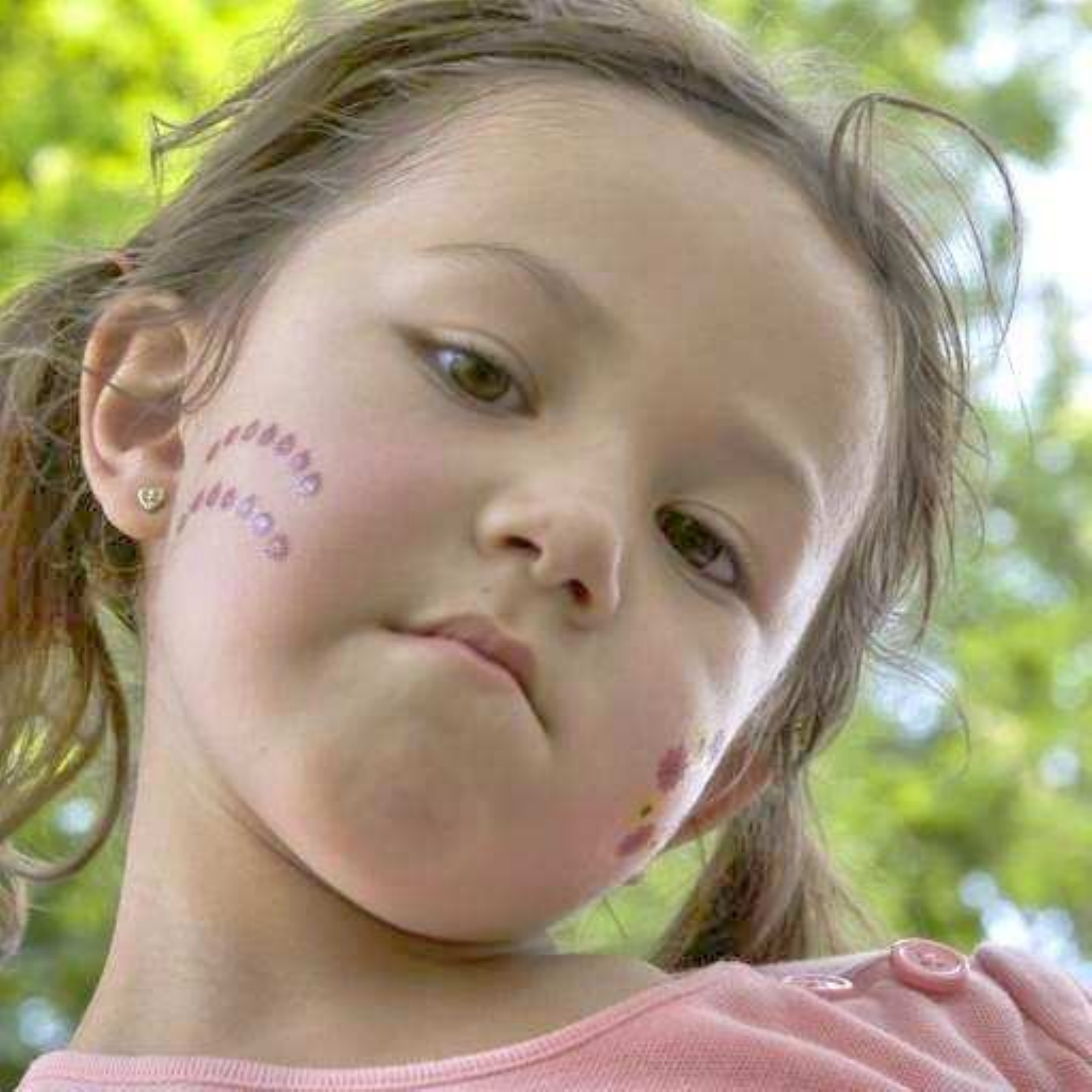}}
    \vskip 0.01truein
    \subcaptionbox
      {WAHE \cite{AriciT2009WAHE}.\label{fig:WAHE5}}
      {\includegraphics[width=0.23\textwidth]{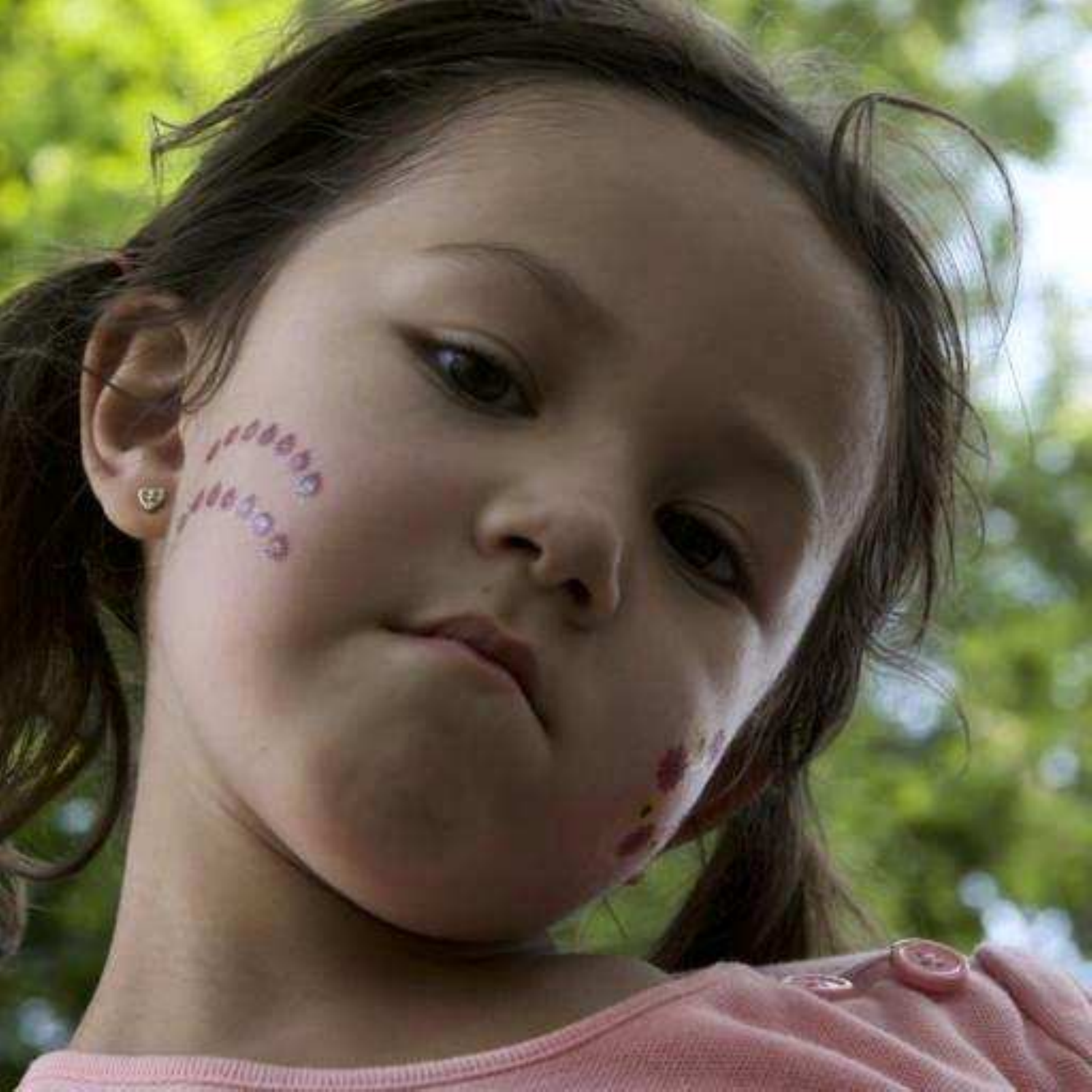}}
    \hskip 0.01truein
    \subcaptionbox
      {DPR \cite{Zhou2020DPR}.\label{fig:DPR5}}
      {\includegraphics[width=0.23\textwidth]{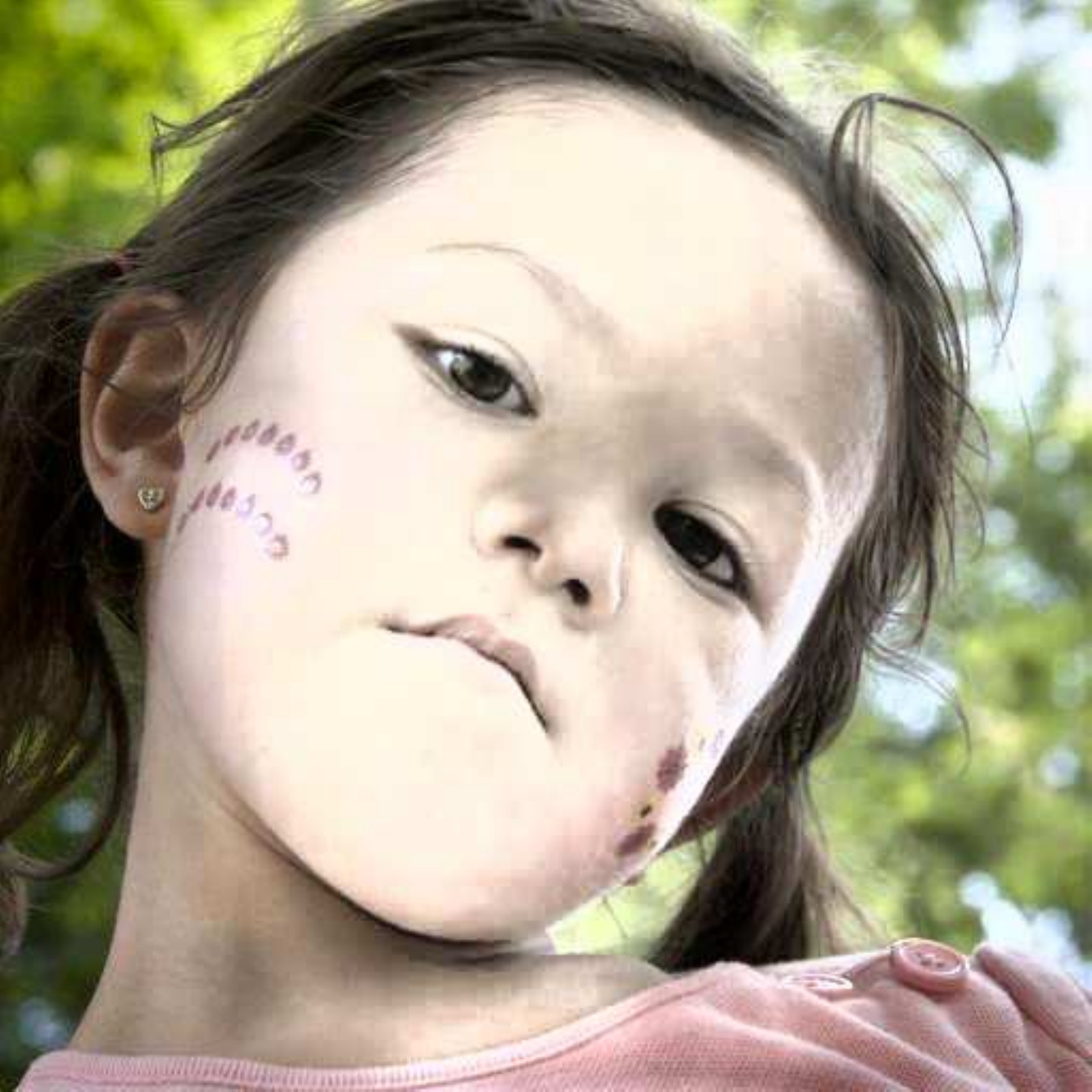}}
    \vskip 0.01truein
    \subcaptionbox
    {PBS \cite{ZhangPBS2021}.\label{fig:PBS5}}
      {\includegraphics[width=0.23\textwidth]{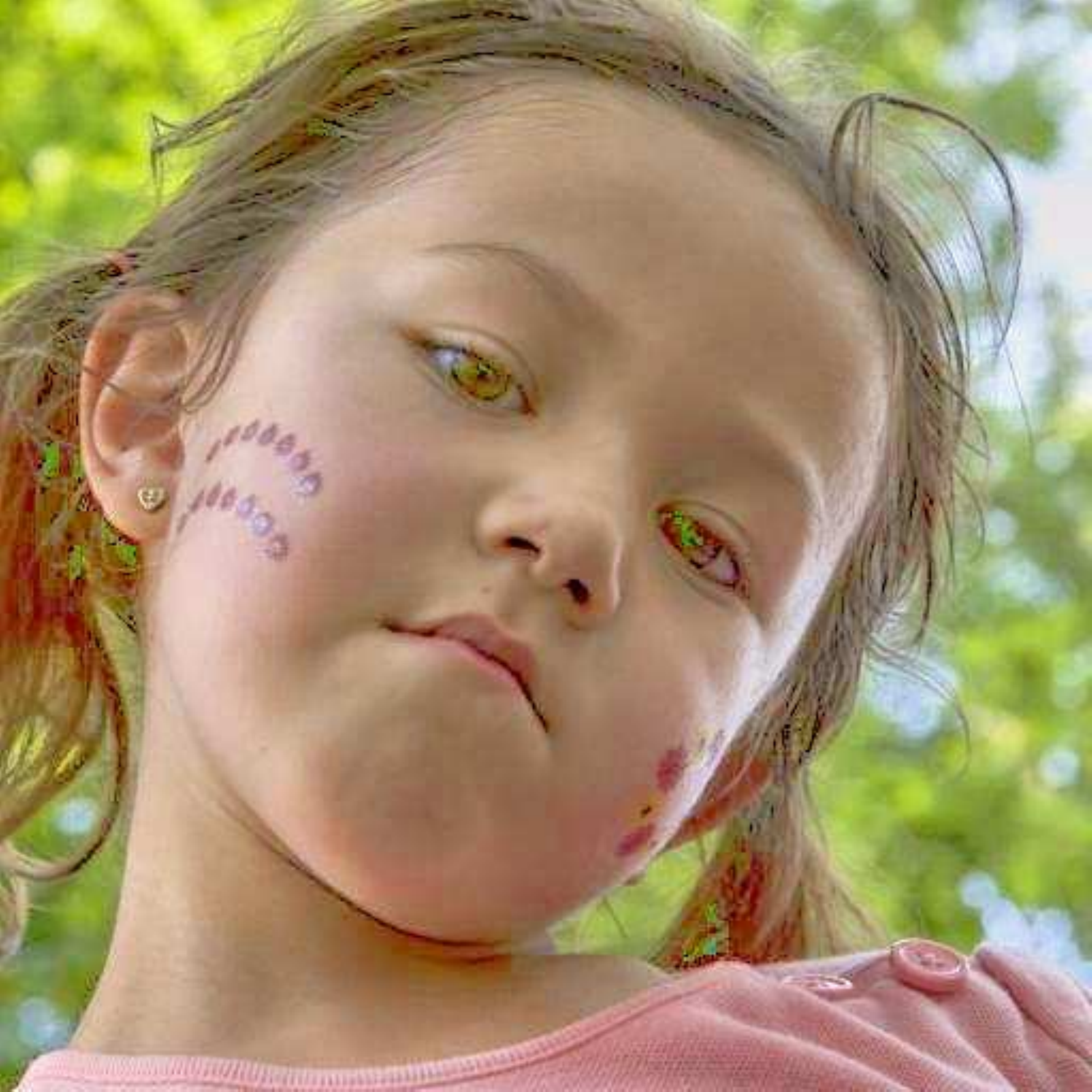}}
    \hskip 0.01truein
    \subcaptionbox
      {Ours.\label{fig:ours5}}
      {\includegraphics[width=0.23\textwidth]{ours_5.pdf}}
    \caption{Qualitative comparisons for different methods (LLCRM, WAHE, DPR, PBS, and ours).}
    \label{fig:comparisons2-5-2}
\end{figure}

First, we would like to compare our results with others qualitatively. The other methods that we choose for this comparison are contrast limited adaptive histogram equalization (CLAHE) \cite{PizerStephenM1987CLAHE}, perceptually bidirectional similarity (PBS) \cite{ZhangPBS2021}, controllable image illumination enhancement (CIIE) \cite{Bai2018Enhance}, structure-revealing low-light (SRLL) enhancement \cite{Li2018SRLL}, simultaneous reflectance and illumination estimation (SRIE) \cite{Fu2016SRIE}, low-light image enhancement using camera reponse model (LLCRM) \cite{Ren2019LLCRM}, weighted adjustable histogram enhancement (WAHE) \cite{AriciT2009WAHE}, and deep portrait relighting (DPR) \cite{Zhou2020DPR}. For all these methods, we use their original code or pretrained model. All these methods are designed for image enhancement except that DPR is designed for face relighting. For the DPR method, we use $pyshtools$ to rotate the original illumination environment to be incident from the top-frontal direction of the subject. This ensures that the pretrained DPR network will receive the best illumination instead of an unbalanced illumination from a side direction. The qualitative results from these methods are shown in Figures \ref{fig:comparisons2-5-1} and \ref{fig:comparisons2-5-2}. 
Our method compares favorably with all eight of the comparsion methods, some of which produce images that still look quite dark, or which have a very unnatural appearance.

\subsection{Quantitative comparisons}
In order to quantitatively compare our results with others, we choose several objective metrics to evaluate the results. Our model is good at balancing the lighting distribution and removing shadows. However, adjusting the relative lightness strength among different areas may risk reversing the lightness order. This may cause unnaturalness and artifacts. So one metric we choose is called the Lightness Order Metric (LOM) \cite{Bai2018Enhance}. It measures the degree of unnaturalness of results by checking if the original lightness order is retained. According to the original paper, smaller scores in the LOM indicate that the lightness order in original images are better retained. And the lighting distribution will be more natural.
For measuring the contrast of images, we use a measure of enhancement (EME) \cite{Agaian2001EME}. For this metric, higher score values mean that the images are better enhanced with higher contrast. 
To quantify the enhancement of details and their visibility, we use the discrete entropy (DE) \cite{Ye2007DE}. Higher DE values represent better enhancement in image details or visibility of images.

Based on our qualitative comparisons, it turns out that results from WAHE and SRIE are poorly enhanced images. The lighting distributions of the enhanced images are almost the same as those of the original inputs. And the underexposed regions are adjusted to be darker. Their enhancement targets are not in accordance with ours. And for DPR, the lighting distributions are always uneven and lots of artifacts are introduced. Based on this, we decided not to include them in the quantitative comparisons. Instead, we will include the remaining methods, which share similar enhancement targets and which yield relatively visually appealing results. So we choose to quantitatively compare our results with LLCRM, SRLL, CIIE, and PBS. The metrics we use are the three metrics we introduced previously. Comparison results using these three metrics are shown in Tables \ref{table:LOM}, \ref{table:EME}, and \ref{table:DE}. We choose nine images from the MIT-Adobe FiveK Dataset as examples. 
For each example, we use bold font to emphasize the method with best performance. For the Lightness Order Metric (LOM), our method gets better lightness naturalness compared with the other methods across all nine images. For the measure of enhancement (EME), our method also gets the highest score for each image. This implies that our enhanced images have good contrast. And for discrete entropy (DE), our method gets the highest average entropy. This shows that the our method enhances details well and increases the visibility of the poorly exposed images.

\begin{table}[h!]
\centering
\begin{tabular}{p{1.3cm}||p{1cm}|p{1cm}|p{1cm}|p{1cm}|p{1cm}}
 \hline
 Methods & LLCRM & SRLL & CIIE & PBS & Ours\\
 \hline
    Image 1 & 0.2141 & 0.2822 & 0.3577 & 0.1494 & \textbf{0.1294} \\
    Image 2 & 0.2500 & 0.3304 & 0.4022 & 0.2237 & \textbf{0.1901} \\
    Image 3 & 0.4888 & 0.3896 & 0.4078 & 0.2772 & \textbf{0.0824} \\
    Image 4 & 0.1867 & 0.5332 & 0.2405 & 0.1536 & \textbf{0.1133} \\
    Image 5 & 0.2338 & 0.2870 & 0.2652 & 0.1872 & \textbf{0.1242} \\
    Image 6 & 0.1420 & 0.2713 & 0.2825 & 0.1186 & \textbf{0.0844} \\
    Image 7 & 0.4319 & 0.3207 & 0.3884 & 0.2171 & \textbf{0.0990} \\
    Image 8 & 0.2149 & 0.2604 & 0.2821 & 0.1564 & \textbf{0.0807} \\
    Image 9 & 0.1415 & 0.4010 & 0.1936 & 0.1446 & \textbf{0.1280} \\
    \hline
    Average & 0.2560 & 0.3418 & 0.3133 & 0.1809 & \textbf{0.1146}\\
 \hline
\end{tabular}
\caption{Comparison results using LOM metric \cite{Bai2018Enhance} (Lower values are better). The images are shown in Fig. \ref{fig:comparisons_montage9}, and are numbered from left to right and top to bottom.}\label{table:LOM}
\end{table}
 
\begin{table}[h!]
\centering
\begin{tabular}{p{1.3cm}||p{1cm}|p{1cm}|p{1cm}|p{1cm}|p{1cm}}
 \hline
 Methods & LLCRM & SRLL & CIIE & PBS & Ours\\
 \hline
    Image 1 & 6.3621 & 5.0363 & 12.7059 & 13.1127 & \textbf{16.2816} \\
    Image 2 & 10.5765 & 8.9197 & 13.4239 & 15.5792 & \textbf{17.1697} \\
    Image 3 & 42.1028 & 23.2063 & 60.6323 & 37.7957 & \textbf{66.1502} \\
    Image 4 & 9.0566 & 10.6031 & 10.2225 & 12.2546 & \textbf{21.8791} \\
    Image 5 & 7.5298 & 5.875 & 9.5969 & 10.5239 & \textbf{11.0064} \\
    Image 6 & 7.0793 & 5.3842 & 7.669 & 8.7756 & \textbf{8.9927} \\
    Image 7 & 13.243 & 9.0524 & 16.7294 & 18.6075 & \textbf{20.6313} \\
    Image 8 & 13.299 & 5.8689 & 25.8228 & 22.8722 & \textbf{30.6223} \\
    Image 9 & 9.6474 & 10.1326 & 11.1961 & 12.8121 & \textbf{14.0283} \\
    \hline
    Average & 13.2107 & 9.3421 & 18.6665 & 16.9259 & \textbf{22.9735}\\
 \hline
\end{tabular}
\caption{Comparison results using EME metric \cite{Agaian2001EME} (Higher values are better). The images are shown in Fig. \ref{fig:comparisons_montage9}, and are numbered from left to right and top to bottom.}\label{table:EME}
\end{table}

\begin{table}[h!]
\centering
\begin{tabular}{p{1.3cm}||p{1cm}|p{1cm}|p{1cm}|p{1cm}|p{1cm}}
 \hline
 Methods & LLCRM & SRLL & CIIE & PBS & Ours\\
 \hline
    Image 1 & 5.1362 & 5.1586 & 4.9258 & 5.2606 & \textbf{5.2641} \\
    Image 2 & 5.0741 & 5.0806 & 4.8626 & 5.1994 & \textbf{5.2537} \\
    Image 3 & 5.0857 & \textbf{5.1163} & 4.8893 & 4.9514 & 4.8512 \\
    Image 4 & 4.9867 & 4.9619 & 5.134 & 5.1718 & \textbf{5.3191} \\
    Image 5 & 5.086 & 5.1171 & 5.0722 & 5.2042 & \textbf{5.257} \\
    Image 6 & 5.0839 & 5.0098 & 4.8257 & 5.1258 & \textbf{5.2468} \\
    Image 7 & \textbf{5.0872} & 5.0599 & 4.9484 & 5.0771 & 5.0088 \\
    Image 8 & 5.045 & 5.0529 & 4.8603 & 5.1529 & \textbf{5.2047} \\
    Image 9 & 4.8471 & 4.8901 & 4.9088 & 5.0302 & \textbf{5.2268} \\
    \hline
    Average & 5.0480 & 5.0497 & 4.9363 & 5.1304 & \textbf{5.1814} \\
 \hline
\end{tabular}
\caption{Comparison results using DE metric \cite{Ye2007DE} (Higher values are better). The images are shown in Fig. \ref{fig:comparisons_montage9}, and are numbered from left to right and top to bottom.}\label{table:DE}
\end{table}

\section{Conclusion} \label{sec:conclude}

We proposed a pipeline to enhance poorly exposed face images. Our method can be separated into two steps: 1) estimating the original lighting distribution, 2) optimizing that lighting distribution. To acquire an accurate lighting distribution, we leveraged ideas from 3D morphable models, face alignment, and face relighting to accurately deduce the face geometry from a single view. 
To further refine the lighting distribution from spherical harmonics, we design our own loss function. We also simplified the function to a convex quadratic form. To validate our method, we compared it both qualitatively and quantitatively with others. The comparison results show that our method performs well and compares favorably with other methods.

{\small
\bibliographystyle{ieee_fullname}
\bibliography{egpaper}
}

\end{document}